Reddened Dimming of Boyajian's Star Supports Internal Storage of Its "Missing" Flux

Peter Foukal, 192 Willow Rd., Nahant, Massachusetts, 01908, USA

The F star KIC 8462852 ("Boyajian's star") has attracted attention by its deep dimming episodes lasting days to weeks (Boyajian et al. 2016) superposed on slower dimming events sometimes lasting for years (Schaefer 2016; Montet & Simon 2016). Originally, it was assumed (e.g. Wright & Sigurdsson 2016) that storage of blocked heat flux inside the star could not explain this behavior, although such internal storage has long been recognized as the mechanism responsible for the photometric variation of magnetically active stars, including the Sun (Spruit 1982; Foukal, Fowler & Livshits 1983).

Recently, we pointed out that such blocking by star-spots might be generalized to other, possibly non-magnetic, obstructions to heat flow in convective stars (Foukal 2017). For example, location of the star near the transition between convective and radiative transport might cause sporadic decreases in heat flux. Alternatively, differential rotation and dynamo action are found to modulate convection in an F star rotating much faster than the Sun (e.g. Augustson, Brun & Toomre 2013).

We show here that such an explanation of the photometric behavior of KIC 8462852 seems to be supported by the star's reddening during two of its sporadic irradiance dips, observed in recent multi-color photometry.

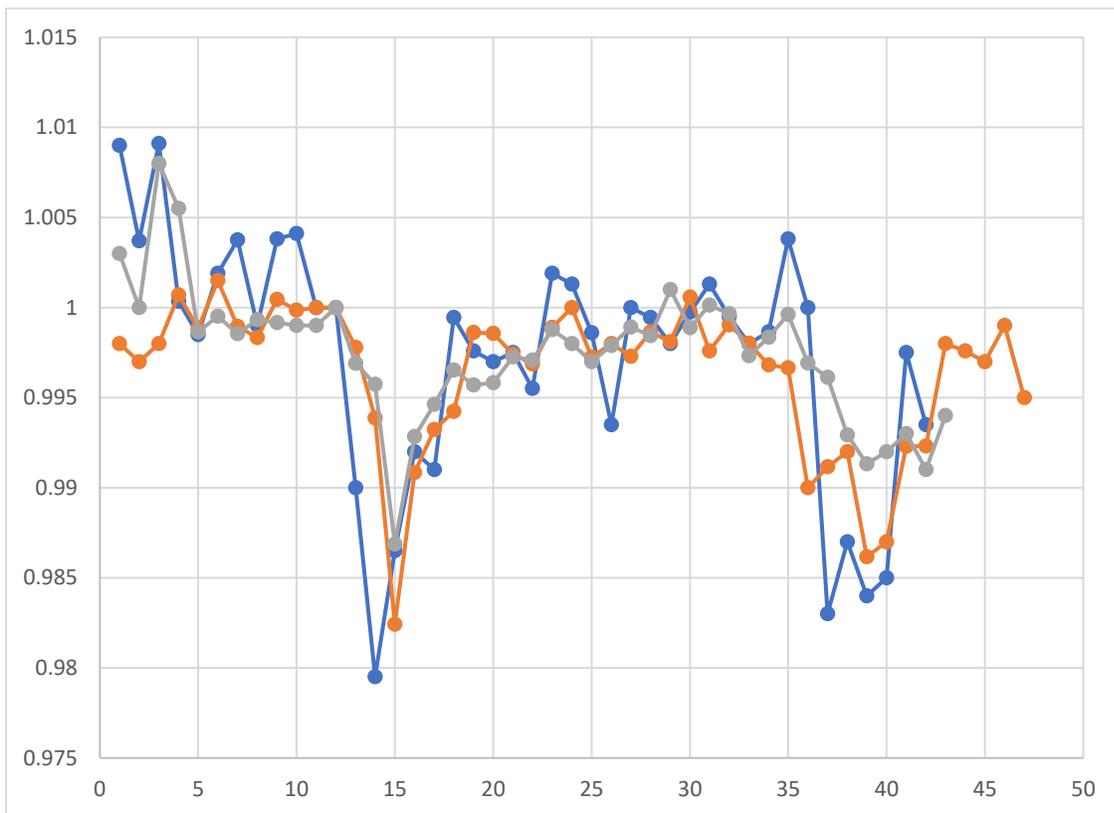

Figure 1. Daily average TFN + OGG photometry between 1 May and 28 June 2017. The abscissa is (Julian date – 2457875). The ordinate is fractional brightness variation. Blue, orange and grey curves show the variation in the B, r' and i' channels respectively.

The three-color photometry made available to us was obtained between 1 May and 28 June 2017 with the 0.4 m telescopes of the Las Cumbres Observatory (LCO) global network on Maui, Hawaii (OGG) and the Canaries, Spain (TFN). The peak transmission wavelengths of the B, r' and i' passbands are 445, 658 and 806 nm. We analyze the daily averages of the corrected data.

We combined the measurements from the TFN and OGG sites and the resultant time series in the three passbands are shown in Fig.1. The main features are two dips whose amplitude decreases from the blue to the IR. Such an inverse relation between amplitude and wavelength is expected for a black body radiator.

Applying the Planck formula with T ~ 6800K for the temperature of this early F star we find a black body cooling of approximately 30 K from the amplitude of the May dip in the blue channel, where the S/N ratio is highest. The amplitudes of the May dip in the B channel relative to the r' and i' channels are then expected to be in the ratios 1.4 and 1.7. This agrees well with the observed ratios 1.4 and 1.6. The less accurate ratios 1.2 and 1.6 observed in the shallower June dip also agree with calculated values to within measurement error. Similar values are obtained when the filter band-passes are integrated over a stellar spectrum, instead of using a black body approximation (Wright 2017).

This agreement with black body behavior is consistent with interpretation of the dips as photospheric cooling. Simon, Shappee, Pojmanski et al. (2017) report reddening during one of the slower dimming events over 6 months, in 2012. Their finding is dependent on less certain inter- calibration of independent data sets. If confirmed it further supports our prediction that the slow dimmings represent release of the blocked flux stored during a dip. Hippke & Angerhausen (2017) report that the slow dimming events do closely follow the sporadic dips.

Reddening of similar magnitude could also be caused by interstellar extinction, but this seems unlikely because the photometric variability of KIC 8462852 requires an improbable geometry of ISM material (e.g. Simon, Shappee & Pojmanski et al 2017). Obscuration by circumstellar grains larger than the wavelength of the observed light (e.g. Meng, Rieke,& Dubois et al. 2017) should not cause reddening.

Our explanation could help to understand the seeming uniqueness of KIC 8462852. Of the roughly 150,000 stars observed by Kepler about 10% are F stars. Fewer than 1% of these might qualify if we require a main sequence early F star at the transition between convective and radiative transport and rotating as fast as KIC 8462852. Such considerations could reduce this star's uniqueness to a less remarkable level closer to 1% of those stars having the required properties.

In conclusion, the most plausible explanation of the reddening of KIC 8462852 during its brief dimmings appears to be photospheric cooling. Together with other evidence on possible reddening of the slower dimmings and on their timing following the brief dips, it favors interpretation as a transient reduction of the star's heat transport efficiency (Foukal 2017; see also Sheikh, Weaver & Dahmen 2016). MHD modelling of heat flow variability in early F stars (e.g. Augustson, Brun & Toomre 2013) could help to identify the specific mechanism of the heat flow blocking.

I thank the LCO staff for obtaining the data and T. Boyajian for making them available.